\documentclass{article}%
\usepackage{amsmath}%
\usepackage{amsfonts}%
\usepackage{amssymb}%
\usepackage{graphicx}
\renewcommand{\not}[1]{#1\!\!\!\slash}

\begin{document}

\title{{\Large \textbf{On the Non-relativistic Limit of Linear Wave Equations for
Zero and Unity Spin Particles}} }
\author{P.Yu. Moshin${}^{a)}$\thanks{Tomsk State Pedagogical University, 634041 Tomsk,
Russia; e-mail: moshin@dfn.if.usp.br}\thinspace\thinspace\thinspace\thinspace
and J.L. Tomazelli${}^{b)}$\thanks{E-mail: tomazelli@fsc.ufsc.br}\\ \\$^{a)}$Instituto de F\'{\i}sica, Universidade de S\~{a}o Paulo,\\Caixa Postal 66318, CEP 05315-970, S\~{a}o Paulo, S.P., Brazil\\$^{b)}$Departamento de F\'{\i}sica, Universidade Federal de Santa Catarina, \\Caixa Postal 476, CEP 88010-970, Florian\'{o}polis, S.C., Brazil}
\date{}
\maketitle

\begin{abstract}
The non-relativistic limit of the linear wave equation for zero and unity spin
bosons of mass $m$ in the Duffin--Kemmer--Petiau representation is
investigated by means of a unitary transformation, analogous to the
Foldy--Wouthuysen canonical transformation for a relativistic electron. The
interacting case is also analyzed, by considering a power series expansion of
the transformed Hamiltonian, thus demonstrating that all features of particle
dynamics can be recovered if corrections of order $1/m^{2}$ are taken into
account through a recursive iteration procedure.

\end{abstract}

\section{Introduction}

Recently, in view of the increasing technical complexity of string theories as
the best candidates for the unification of the fundamental interactions, there
is a renewed interest in the quantum field theory of higher spins as a natural
covariant formalism for accommodating the particle spectra in the Standard
Model and quantum gravity theories, as well as in their supersymmetric
counterparts. Thus, from the phenomenological standpoint, it is mandatory to
investigate such theories in the low-energy regime, by examining their
non-relativistic formal properties and taking into account the interaction
with external electromagnetic and/or metric fields as a starting point.

In relativistic quantum mechanics, one must seek for a relation between
irreducible representations of the Poincar\'{e} group and wave equations. In
Wigner's standard form, non-trivial wave equations can only be presented for
wave functions with a large number of components, simultaneously expressing
constraints on redundant components and equations of motion for the physical
ones. Considering general invariant equations, Gel'fand and
Yaglom$^{{\scriptsize \cite{r1}}}$ expressed relativistic wave functions in
terms of linear differential operators, simultaneously determining both these
operators and the finite-dimensional representations of the homogeneous
Lorentz group, according to which the components of the wave functions
transform. However, such a procedure is not applicable to non-relativistic
wave equations whose solutions transform according to the homogeneous Galilei
group. Following another approach, relying upon the Bargmann--Wigner method,
L\'{e}vi-Leblond$^{{\scriptsize \cite{r2}}}$ constructed a basis in a
ten-dimensional representation space of the homogeneous Galilei group for free
massive particles of spin $1$, by taking a complete set of independent linear
combinations of symmetrical tensor products of two-component wave functions
which describe non-relativistic particles of spin $1/2$, and arriving at a
system of equations involving linear operators.

In order to investigate the physical properties of particles of zero and unity
spin in the presence of electromagnetic external sources, instead of starting
from Galilean-covariant wave equations, we start from a Lorentz-covariant
linear wave equation in the Hamiltonian form and apply a canonical
transformation, analogous to the Foldy--Wouthuysen (FW)
transformation$^{{\scriptsize \cite{r3}}}$ for Dirac fermions, to a suitable
reference frame in which one can recognize the different couplings of charged
bosons with the electromagnetic field. In this sense, the covariant linear
representation$^{{\scriptsize \cite{r4,r42,r43}}}$ of Duffin--Kemmer--Petiau
(DKP) proves to be particularly useful, since all physical quantities are
constructed from linear operators which obey convenient algebraic relations,
in close similarity with the familiar Dirac operators. Notably,
Darwin$^{{\scriptsize \cite{da}}}$ proposed a linear wave equation for the
electromagnetic field some years before Petiau's pioneering work, in close
relation to the meson theory proposed by Kemmer, who referred to Dirac's
work$^{{\scriptsize \cite{di}}}$ on linear relativistic equations for
particles with spins higher than one-half.

This work is organized as follows. In Section 2, we present the linear wave
equation which describes bosons of spin zero and unity and the basic
identities of the associated DKP algebra; we then rewrite this equation in the
Hamiltonian form for non-interacting particles. In Section 3, we discuss the
quantum canonical transformation for the free boson Hamiltonian, by analogy
with the ordinary FW transformation. Next, in Section 4, we derive the
non-relativistic limit of the Hamiltonian that describes charged bosons
interacting with an external electromagnetic field. In Section 5, we make
concluding remarks.

\section{DKP Hamiltonian}

Let us briefly review the DKP formalism for non-interacting bosons of spin
zero and one. The relativistic wave equation in such a representation reads
\begin{equation}
\left(  i\not \partial -m\right)  \psi=0\,,\label{1}%
\end{equation}
where $\not \partial \equiv\beta_{\mu}\partial^{\mu}$ and $\psi$ is a
five(ten)-row column associated with the zero (unity) spin field. The
considerations of this work do not refer to any particular representation for
$\psi$.

The $\beta$-matrices obey the algebra
\begin{equation}
\beta_{\mu}\beta_{\nu}\beta_{\rho}+\beta_{\rho}\beta_{\nu}\beta_{\mu}%
=\beta_{\mu}g_{\nu\rho}+\beta_{\rho}g_{\nu\mu}\,,\label{2}%
\end{equation}
which implies the following consequences:%
\begin{align}
&  \,\beta_{0}\beta_{k}\beta_{0}=0\,,\;k=1,2,3\,,\nonumber\\
&  \,\beta_{0}^{3}=\beta_{0}\,,\label{4}\\
&  \,\not b  \beta_{\nu}\not b  =\not b  b_{\nu}\,,\label{5}\\
&  \,(\vec{\beta}\cdot\vec{b})\beta_{0}(\vec{\beta}\cdot\vec{b})=0\,,\label{6}%
\end{align}
where $b_{\mu}=(b_{0},\vec{b})$ is a generic four-vector.

Multiplying (\ref{1}) by $\not \partial \beta_{\mu}$ and using (\ref{5}),
\[
\left(  i\not \partial \beta_{\mu}\not \partial -m\not \partial \beta_{\mu
}\right)  \psi=\left(  i\partial_{\mu}\not \partial -m\not \partial \beta
_{\mu}\right)  \psi=0\,,
\]
and then (\ref{1}),
\[
\left(  m\partial_{\mu}-m\not \partial \beta_{\mu}\right)  \psi=0\,,
\]
one obtains%
\begin{equation}
\partial_{\mu}\psi=\not \partial \beta_{\mu}\psi\,.\label{7}%
\end{equation}
Multiplying (\ref{1}) by $\beta_{0}$ and taking the zero component of
(\ref{7}), times the imaginary unity, one obtains, upon adding the results,
\[
\left\{  i\left[  \partial_{0}+\partial^{k}\left(  \beta_{0}\beta_{k}%
-\beta_{k}\beta_{0}\right)  \right]  -m\beta_{0}\right\}  \psi=0\,,
\]
or
\begin{equation}
i\partial_{0}\psi=H\psi\,,\label{8}%
\end{equation}
where
\begin{equation}
H=-i\vec{\alpha}\cdot\vec{\nabla}+\beta_{0}m=\vec{\alpha}\cdot\vec{p}%
+\beta_{0}m\label{9}%
\end{equation}
is the DKP Hamiltonian, and $\vec{\alpha}$ is defined by its spatial
components:%
\[
\alpha_{k}\equiv\beta_{0}\beta_{k}-\beta_{k}\beta_{0}\,,\;k=1,2,3\,.
\]

\section{FW Transformation}

As in the electron case, we now look for a unitary transformation,%
\begin{align*}
\psi^{\prime} &  =e^{iU}\psi\,,\\
H^{\prime} &  =e^{iU}He^{-iU}\,,
\end{align*}
which should eliminate the term that involves the spatial components of the
four-momentum. In case $H$ explicitly depends on time, equation (\ref{8})
yields
\[
i\partial_{0}(e^{-iU}\psi^{\prime})=He^{-iU}\psi^{\prime}\,,
\]
so that
\[
e^{-iU}\left(  i\partial_{0}\psi^{\prime}\right)  =\left(  He^{-iU}%
-i\partial_{0}e^{-iU}\right)  \psi^{\prime}\,,
\]
or
\[
i\partial_{0}\psi^{\prime}=H^{\prime}\psi^{\prime}\,,
\]
where
\begin{equation}
H^{\prime}=e^{iU}\left(  H-i\partial_{0}\right)  e^{-iU}\,.\label{14}%
\end{equation}

Let us choose
\[
U=-i\frac{\vec{\beta}.\vec{p}}{|\vec{p}|}\theta\,.
\]
The $\beta$-algebra (\ref{2}) implies the identity%
\begin{align}
2(\vec{\beta}.\vec{p})^{3}  &  =\sum_{ijk}p_{i}p_{j}p_{k}\left(  \beta
_{i}\beta_{j}\beta_{k}+\beta_{k}\beta_{j}\beta_{i}\right) \nonumber\\
&  =-\sum_{ijk}p_{i}p_{j}p_{k}(\beta_{i}\delta_{jk}+\beta_{k}\delta
_{ji})\,,\nonumber
\end{align}
so that%
\begin{equation}
(\vec{\beta}\cdot\vec{p})^{3}=-|\vec{p}|^{2}(\vec{\beta}\cdot\vec
{p})\,.\label{16}%
\end{equation}
Representing (\ref{16}) in the form
\[
\lbrack(\vec{\beta}\cdot\vec{p})^{2}+\left|  \vec{p}\right|  ^{2}](\vec{\beta
}\cdot\vec{p})=0
\]
and then, on the mass shell,%
\[
(\vec{\beta}\cdot\vec{p})=\beta_{0}p_{0}-\not p =\beta_{0}E+m\,,
\]
we have
\begin{equation}
\lbrack(\vec{\beta}\cdot\vec{p})^{2}+\left|  \vec{p}\right|  ^{2}]\left(
\beta_{0}E+m\right)  \psi=0\,.\label{17}%
\end{equation}
On the other hand, due to the identity%
\begin{align}
(\vec{\beta}\cdot\vec{p})^{2}\beta_{0}  &  =\sum_{ij}p_{i}p_{j}\left(
\beta_{i}\beta_{j}\beta_{0}\right) \nonumber\\
&  =-\sum_{ij}p_{i}p_{j}\left(  \beta_{0}\beta_{j}\beta_{i}+\beta_{0}%
\delta_{ij}\right)  =-\beta_{0}[(\vec{\beta}\cdot\vec{p})^{2}+|p|^{2}%
]\,,\nonumber
\end{align}
equation (\ref{17}) implies
\[
\left(  m-\beta_{0}E\right)  (\vec{\beta}\cdot\vec{p})^{2}\psi=-\left|
\vec{p}\right|  ^{2}m\psi\,,
\]
or
\[
(m^{2}-\beta_{0}^{2}E^{2})(\vec{\beta}\cdot\vec{p})^{2}\psi=-(m^{2}+\beta
_{0}Em)\left|  \vec{p}\right|  ^{2}\psi\,.
\]
Then, due to (\ref{4}), we obtain
\begin{equation}
(\vec{\beta}\cdot\vec{p})^{2}=-\left|  \vec{p}\right|  ^{2}+\left(  Em\right)
\beta_{0}+E^{2}\beta_{0}^{2}\,.\label{19}%
\end{equation}
Eq. (\ref{9}) does not contain the complete information about the system
because of multiplication by the singular matrix $\beta_{0}$.

Multiplying (\ref{1}) by $(1-\beta_{0}^{2})$, one finds the additional
constraint
\[
\left[  i\partial^{k}\beta_{k}\beta_{0}^{2}-(1-\beta_{0}^{2})m\right]
\psi=0\,,
\]
or
\begin{equation}
(\vec{\beta}\cdot\vec{p})\beta_{0}^{2}+(1-\beta_{0}^{2})m=0\,,\label{20}%
\end{equation}
on the mass shell. Also, left-multiplying (\ref{19}) by $(\vec{\beta}\cdot
\vec{p})$, and using (\ref{16}) and (\ref{20}), one obtains%
\begin{equation}
(\vec{\beta}\cdot\vec{p})\beta_{0}=E(1-\beta_{0}^{2})\,.\label{21}%
\end{equation}
Now, multiplying (\ref{19}) by $(\vec{\beta}\cdot\vec{p})^{2}$ and using
(\ref{20}), (\ref{21}), one gets%
\begin{equation}
(\vec{\beta}\cdot\vec{p})^{4}=-(\vec{\beta}\cdot\vec{p})^{2}\left|  \vec
{p}\right|  ^{2}\,.\label{22}%
\end{equation}
Then
\[
e^{iU}=e^{(\vec{\beta}\cdot\vec{p}/|\vec{p}|)\theta}=1+\frac{(\vec{\beta}%
\cdot\vec{p})^{2}}{|\vec{p}|^{2}}\left(  1-\cos\theta\right)  +\frac{(\vec
{\beta}\cdot\vec{p})}{|\vec{p}|}\sin\theta\,,
\]
where (\ref{16}) and (\ref{22}) have been used in the series expansion.
Hence,
\[
H^{\prime}=(\vec{\alpha}.\vec{p})\left(  \cos\theta-\frac{m}{|\vec{p}|}%
\sin\theta\right)  +\beta_{0}\left(  |\vec{p}|\sin\theta+m\cos\theta\right)
\,.
\]
Choosing
\[
\sin\theta=\frac{|\vec{p}|}{E}\,,\;\cos\theta=\frac{m}{E}\,,
\]
one arrives at%
\[
H^{\prime}=\frac{\beta_{0}}{E}\left(  \vec{p}^{2}+m^{2}\right)  =\beta_{0}E\,.
\]

\section{DKP Interaction Hamiltonian}

In order to have a better understanding of the particle content of the theory,
let us examine the behavior of charged bosons in the presence of an external
electromagnetic field, transformed to a reference frame where particles carry
low momenta. The electromagnetic interaction is introduced by means of the
covariant derivative, so that%
\begin{equation}
(i\not D  -m)\psi=0\,,\label{27}%
\end{equation}
where the covariant derivative%
\[
D_{\mu}=\partial_{\mu}+ieA_{\mu}%
\]
satisfies the commutation relation%
\[
\left[  D_{\mu},D_{\nu}\right]  =ieF_{\mu\nu}\,.
\]

Multiplying (\ref{27}) by $\not D\beta_{\mu}$, one
obtains,$^{{\scriptsize \cite{r43}}}$ by analogy with (\ref{7}),
\begin{equation}
D_{\mu}\psi=\not D\beta_{\mu}\psi+\frac{e}{2m}F^{\rho\sigma}(\beta_{\rho}%
\beta_{\mu}\beta_{\sigma}-\beta_{\rho}g_{\mu\sigma})\psi\,.\label{30}%
\end{equation}
Then, from equations (\ref{27}) and (\ref{30}),%
\[
i\partial_{0}\psi=H\psi\,,
\]
it follows that%
\[
H=H^{(0)}+H^{(1)}\,,
\]
where
\begin{align*}
&  H^{(0)}=\vec{\alpha}\cdot\vec{\pi}+m\beta_{0}-eA_{0}\,,\\
&  H^{(1)}=\frac{ie}{2m}F^{\rho\sigma}\left(  \beta_{\rho}\beta_{0}%
\beta_{\sigma}-\beta_{\rho}g_{0\sigma}\right)  \,,
\end{align*}
and
\[
\vec{\pi}=\vec{p}-e\vec{A}\,.
\]

Using (\ref{14}) and the Baker--Campbell--Hausdorf formula, one can
write$^{{\scriptsize \cite{r3}}}$%
\[
H{}^{\prime}=H+\frac{\partial U}{\partial t}+i\left[  U,H+\frac{1}%
{2}\frac{\partial U}{\partial t}\right]  -\frac{1}{2!}\left[  U,\left[
U,H+\frac{1}{3}\frac{\partial U}{\partial t}\right]  \right]  +\dots\,.
\]
By virtue of the nonrelativistic limit $\theta\sim\sin\theta\sim|\vec{p}|/m$,
one can choose, in the first approximation, by analogy with the free case,%
\[
U=-i\frac{\vec{\beta}{\cdot}\vec{\pi}}{m}\,.
\]
From the commutation relations (\ref{A1})--(\ref{A7}) and the vector
identities (\ref{B1}), (\ref{B2}), listed in the Appendix, one obtains
\[
\lbrack U,H^{(1)}]=-\frac{e}{m^{2}}[\vec{\beta}\cdot\vec{\pi},(\vec{\beta
}\cdot\vec{E})\beta_{0}^{2}]+\frac{e}{m^{2}}[\vec{\beta}\cdot\vec{\pi}%
,\vec{\beta}\cdot\vec{E}]-\frac{e}{2m^{2}}[\vec{\beta}\cdot\vec{\pi}%
,F^{ij}\beta_{i}\beta_{0}\beta_{j}]\,.
\]
In addition,%
\[
\lbrack\vec{\beta}\cdot\vec{\pi},(\vec{\beta}\cdot\vec{E})\beta_{0}^{2}%
]=i\vec{S}\cdot\lbrack\vec{\pi}\times\vec{E}]\beta_{0}^{2}+(\vec{\beta}%
\cdot\vec{E})[2(\vec{\beta}\cdot\vec{\pi})\beta_{0}^{2}-\vec{\beta}\cdot
\vec{\pi}]\,,
\]
so that one arrives at%
\begin{align}
H{}^{\prime} &  =m\beta_{0}-eA_{0}+\frac{{\vec{\pi}}^{2}}{2m}\left(
\vec{\alpha}\cdot\frac{\vec{\pi}}{m}-\beta_{0}\right)  +\frac{e}{2m}(\vec
{S}\cdot\vec{H})\beta_{0}+\frac{e}{2m}(\vec{\beta}\times\vec{\alpha})\cdot
\vec{H}\nonumber\\
&  +\frac{e}{2m^{2}}(\vec{S}\cdot(\vec{\pi}\times\vec{E}))\left(  1+2\beta
_{0}^{2}\right)  +\frac{ie}{2m^{2}}[\vec{\beta}\cdot\vec{\pi},(\beta_{0}%
\vec{S}+\vec{\beta}\times\vec{\alpha})\cdot\vec{H}]\nonumber\\
&  -\frac{ie}{m}(\vec{\beta}\cdot\vec{E})\beta_{0}^{2}-\frac{ie}{m^{2}}%
(\vec{\beta}\cdot\vec{E})[2(\vec{\beta}\cdot\vec{\pi})\beta_{0}^{2}-\vec
{\beta}\cdot\vec{\pi}]+\mathcal{O}\left(  m^{-3}\right)  ,\label{40}%
\end{align}
where relation (\ref{B3}) has been used. In the above expression, $\vec{S}$
corresponds to the spin operator of bosons,%
\[
S_{ij}=i\left(  \beta_{i}\beta_{0}\beta_{j}-\beta_{j}\beta_{0}\beta
_{i}\right)  \,,\;i,j=1,2,3\,,
\]
with eigenvalues $0$ or $1$, while $\vec{E}$ and $\vec{H}$ are the electric
and magnetic fields, respectively.

Expression (\ref{40}) is analogous to the Hamiltonian of the Pauli equation
for spin-$1/2$ fermions in the case of charged bosons of spin $0$ and $1$ on
the background of an external electromagnetic field. In (\ref{40}), we can
recognize each term individually. For example, the second term is related to
the electrostatic potential, while the third one corresponds to the kinetic
term of the non-relativistic interaction Hamiltonian. In fact, taking the same
steps that led to equation (\ref{21}) on the mass shell, and using the
definition of the matrices $\alpha_{k}$, one can rewrite the kinetic term in
the transformed Hamiltonian as the expression%
\[
\frac{{\vec{\pi}}^{2}}{2m}\left[  \frac{\pi_{0}}{m}(2\beta_{0}^{2}-1)\right]
\,,
\]
which is indeed diagonal and non-singular in the matrix realization of the DKP
$\beta$-algebra, as one should expect by analogy with the disentangling
property of the FW transformation.

In this approach, the most essential result is the appearance of the spin and
orbital angular momentum couplings with the external magnetic field (the
fourth and fifth terms, respectively), as well as the diagonal spin-orbital
coupling (the sixth term) via the electric field; the last two terms may be
interpreted as being similar to the Darwin term for spin-$1/2$ fermions in the
presence of an electric field; the remaining terms represent higher-order
corrections to such effects, as well as the (non-diagonal) corrections to the
rest-energy (the first term).

\section{Concluding Remarks}

In the preceding sections, we have investigated the non-relativistic limit of
the Lorentz-invariant wave equation which describes scalar and vector mesons
in the so-called Duffin--Kemmer--Petiau representation. By constructing
unitary operators involving the spatial components of the relativistic
$4$-momentum and those belonging to the associated DKP algebra, both for free
particles and for charged bosons in an electromagnetic background, we
performed a quantum canonical transformation to a reference frame where we
succeeded in identifying the coupling terms with the electric and magnetic
fields, in close similarity with the non-relativistic behaviour of interacting
fermions described by the Pauli equation.

Our approach differs from that of L\'{e}vi-Leblond$^{{\scriptsize \cite{r2}}}
$ in the sense that he derived non-relativistic linear wave equations for
particles of arbitrary spins which obey the Galilean invariance by
construction, where the electromagnetic multipole moments are introduced on
dimensional grounds. At the same time, in the case of massive particles of
spin $1$ he settled the corresponding wave equations by employing the
Bargmann--Wigner construction, not referring to the algebraic properties of
the quantities involved, which we have done explicitly in our treatment.

In the context of the present work, it is relevant to mention the series of
papers \cite{r51,r52,r53,r54} by Fushchych, Nikitin et al., who first
introduced non-relativistic Duffin--Kemmer--Petiau equations: in \cite{r51},
the authors presented a discussion on the Galilean-invariant equations for
free particles with arbitrary spins, with particular emphasis on spin $0$ and
$1/2$; however, in the presence of external fields \cite{r52,r53} their
transformation operator, which partially diagonalizes the total Hamiltonian,
differs from that of Foldy and Wouthuysen for spin $1/2$ particles, as they
pointed out in \cite{r54}. Consequently, the transformation operator of
\cite{r52} is not the same as ours, since we follow the steps of the
Foldy--Wouthuysen original algebraic construction (see, e.g., eqs. (5.2),
(5.12)--(5.14) of \cite{r53}), instead of appealing to formal group
theoretical reasonings.

Yet in the framework of the DKP theory, other authors have recently
investigated the non-relativistic wave equation for spinless bosons, also via
the Galilean covariance, by introducing an extra degree of freedom into the
free Lagrangian density$^{{\scriptsize \cite{r6}}}$, thus recovering the
Schr\"{o}dinger equation for a free particle. However, the introduction of
electromagnetic potentials spoils the original structure of the associated Lie
algebra on which the reasoning$^{{\scriptsize \cite{r6}}}$ is grounded.

An interesting issue related to the present work is a possible generalization
of the above procedure to theories of higher spins, as well as to their
non-Abelian counterparts$^{{\scriptsize \cite{r7}}}$.

\textbf{Acknowledgments} The authors are grateful to D.M. Gitman for useful
discussions. The work was supported by CNPq.

\appendix

\section{Appendix}

\setcounter{equation}{0}  \renewcommand{\theequation}{A.\arabic{equation}}

Below, we present some useful commutation and vector relations derived from
the algebra (\ref{2}) of the $\beta$-matrices:%
\begin{align}
&  \left[  {U,\vec{\alpha}\cdot\vec{\pi}}\right]  {=\frac{i}{m}\beta_{0}%
\vec{\pi}^{2}}\,\,,\label{A1}\\
&  \left[  {U,\beta_{0}}\right]  {=\frac{i}{m}\vec{\alpha}\cdot\vec{\pi}%
}\,\,,\label{A2}\\
&  \left[  {U,A_{0}}\right]  {=-\frac{i}{m}\vec{\beta}\cdot\vec{\nabla}A_{0}%
}\,,\label{A3}\\
&  {\left[  U,\partial U/\partial t\right]  =\frac{ie}{m^{2}}\vec{S}%
\cdot\left(  \vec{\pi}\times\partial\vec{A}/\partial t\right)  }%
\,,\label{A4}\\
&  \left[  {U,}\left[  {U,\vec{\alpha}\cdot\vec{\pi}}\right]  \right]
{=-\frac{\vec{\pi}^{2}}{m^{2}}}\left(  {\vec{\alpha}\cdot\vec{\pi}}\right)
\,,\label{A5}\\
&  \left[  {U,}\left[  {U,\beta_{0}}\right]  \right]  {=-\frac{1}{m^{2}}%
\beta_{0}\vec{\pi}^{2}}\,,\label{A6}\\
&  \left[  {U,}\left[  {U,A_{0}}\right]  \right]  {=-\frac{1}{m^{2}}\vec
{S}\cdot}({\vec{\pi}\times\vec{\nabla}A_{0})}\,,\label{A7}\\
&  {F^{\rho\sigma}\beta_{\rho}\beta_{0}\beta_{\sigma}=-2}({\vec{E}\cdot
\vec{\beta})\beta_{0}^{2}+\vec{E}\cdot\vec{\beta}+F^{ij}\beta_{i}\beta
_{0}\beta_{j}}\,,\label{B1}\\
&  {F^{\rho\sigma}\beta_{\rho}g_{0\sigma}=-\vec{E}\cdot\vec{\beta}%
}\,,\label{B2}\\
&  {F^{ij}\beta_{i}\beta_{0}\beta_{j}=-i\beta_{0}\vec{S}\cdot\vec{H}-i}%
({\vec{\beta}\times\vec{\alpha})\cdot\vec{H}}\,.\label{B3}%
\end{align}

\end{document}